\documentclass[11pt]{article}

\usepackage[preprint]{acl}

\usepackage{times}
\usepackage{latexsym}

\usepackage[T1]{fontenc}

\usepackage[utf8]{inputenc}

\usepackage{microtype}

\usepackage{inconsolata}

\usepackage{graphicx}
\usepackage{subfig}
\usepackage{booktabs}
\usepackage{graphicx}
\usepackage{amsmath}
\usepackage{amssymb}
\usepackage{enumitem}
\usepackage{multirow}
\usepackage{diagbox}
\usepackage[table]{xcolor}
\usepackage{fontawesome5}   
\usepackage{hyperref}       
\title{Understanding the Self-Reflection Mechanisms of LLMs through \\
Biased Attitude Associations}

\author{Jingshen Zhang$^{1}$, Bo Wang$^{1{\dagger}}$, Boci Yang$^{2}$, Dongming Zhao$^1$, Ruifang He$^1$, \\
\textbf{Yuexian Hou$^1$, Zifei Yu$^3$} \\
  $^1$ School of Computer Science and Technology, Tianjin University \\
  $^2$ School of New Media and Communication, Tianjin University \\
  $^3$ Tianjin Huizhixingyuan Information Technology Co., Ltd \\
  \texttt{\{jason\_zhang, bo\_wang\}@tju.edu.cn} 
  \\}

\makeatletter
\newcommand\blfootnote[1]{%
  \begingroup
  \renewcommand\thefootnote{}\footnotetext{#1}%
  \addtocounter{footnote}{-1}%
  \endgroup
}
\makeatother

\begin{document}
\maketitle 
\blfootnote{$^\dagger$ Corresponding author.} 

\begin{abstract}
While the emergent self-reflection capabilities of Large Language Models (LLMs) offer a promising paradigm for autonomous bias mitigation, their internal mechanics remain unclear, raising concerns regarding potential bias entrenchment. Under the premise that social bias is intrinsically encoded as \textit{valence} inclinations, where the exacerbation of bias scales with sharper valence fluctuations across social groups, this paper proposes \textit{ReBias-Lens}, a probing framework designed to interpret how self-reflection reconfigures these biased attitude associations through the lens of valence projection within \textit{intersectional} contexts. Central to ReBias-Lens is the metric of \textit{Valence Fluctuation (VF)} comprising two variants: \textit{Global-VF}, which captures macroscopic valence encoding trends, and \textit{Local-VF}, which scrutinizes microscopic distinctiveness across specific social categories. Deploying ReBias-Lens to evaluate four LLMs across twelve social categories reveals that overall valence fluctuations undergo a distinct layer-wise smoothing, characterized by a significant hierarchical representation divergence as the layers deepen, which ultimately manifests as a widespread mitigation of bias at the behavioral level. In stark contrast to this macro-level reduction, this reflection mechanism is not universally corrective, instead exhibiting a stubborn, category-specific selectivity that regularly locks in and perversely amplifies localized biases. \textcolor{red}{Warning: this paper contains examples with biased content.}

\href{https://github.com/JasonZhang0702/rebias-lens}{%
  \faGithub\hspace{0.3em}github.com/JasonZhang0702/rebias-lens
}
\end{abstract}

\section{Introduction}

Leveraging superior in-context learning, Large Language Models (LLMs) have demonstrated surprising emergent self-reflection capabilities, offering a novel paradigm for autonomous decision-making and safety governance \cite{shinn2023reflexionlanguageagentsverbal, ji-etal-2023-towards, Gallego2024MergingIS, phan-etal-2025-think, zhao-etal-2025-explicit,liu2025selfreflectionmakeslargelanguage}. This potential is especially pronounced in algorithmic fairness, where self-reflection mechanisms offer a promising avenue to mitigate stereotypically biased content \cite{ganguli2023capacitymoralselfcorrectionlarge,cheng2024reinforcementlearningmultiroledebates,borah2024implicitbiasdetectionmitigation,lou2024anchoringbiaslargelanguage, zhao-etal-2025-explicit, gallegos-etal-2025-self}. Yet, recent evidence suggests that self-reflection is not inherently synonymous with neutrality; conversely, it may precipitate self-preference amplification and systemic overconfidence \cite{xu2024prideprejudicellmamplifies, ren2024biasamplificationlanguagemodel, zhang-etal-2025-safeconf, zhang2025understandingdarkllmsintrinsic}. Confronting these latent vulnerabilities, this study argues that the advancement of \textit{self-reflection necessitates interpretability} interventions to expose and preemptively disrupt bias entrenchment within LLMs, thereby paving the way for more reliable autonomous agent self-evolution.

Previous works have systematically evaluated the prerequisites for effective self-reflection from a behavioral perspective \cite{kamoi-etal-2024-llms}, yet failed to investigate the underlying mechanisms. Concurrently, other studies have employed probing techniques to trace the inter-layer propagation of toxicity during self-reflection \cite{Liu2024SelfcorrectionIN, liu-etal-2024-intrinsic}, yet they rely on explicit toxicity metrics that struggle to capture the implicit and nuanced value inclinations inherent in social biases. These collective limitations underscore a critical research lacuna: \textit{the lack of an internal, granular lens} capable of deciphering how LLMs process subtle, implicit biases during self-reflection. 

To bridge this gap, this paper proposes \textit{\textbf{ReBias-Lens}, an interpretability framework that integrates valence projection technology within intersectional contexts.} Specifically, it operationalizes social biases as measurable shifts along the internal valence spectrum, utilizing intersectional prompts as a diagnostic lens to expose implicit prejudices. Consequently, this framework enables a multidimensional and dynamic investigation of how self-reflection reconfigures biased attitude associations. Central to {ReBias-Lens} is the quantification of \textbf{\textit{Valence Fluctuation (VF)}} comprising two variants: \textbf{\textit{Global-VF}}, which captures macroscopic valence encoding trends, and \textbf{\textit{Local-VF}}, which scrutinizes microscopic distinctiveness across specific social categories (Section \ref{sec:rebias-lens-framework}). By decoupling these fluctuations, ReBias-Lens provides a granular lens to dissect the mechanics of self-reflection. We deploy this framework across four representative LLMs and twelve social categories. Our analysis is guided by following three hierarchically organized research questions:

\paragraph{\textbf{RQ1:}} \textit{To what extent does self-reflection systematically modulate the magnitude of bias via cross-sample VF?} This question seeks to provide a macroscopic account of how self-reflection restructures biased attitude associations, examining whether it mitigates or amplifies bias and how these impacts are expressed in behavioral outputs, thus supporting mechanistic analyses in subsequent research questions. Experimental results demonstrate that self-reflection drives stable valence smoothing, with Global-VF reduced by more than 73\% on average across models and templates. Moreover, our framework not only clarifies the mechanics of LLMs’ normative behaviors but also offers a new tool for analyzing exaggerated safety phenomena (Section \ref{sec:rq1}).
\paragraph{\textbf{RQ2:}} \textit{Do observed smoothing effects of VF originate from specific neural layers?} This question is motivated by our observation that VF displays distinctive hierarchical smoothing across model layers. To answer it, we first employ a change-point detection algorithm to identify the key transition layer $l^*$, then conduct layer-wise window-based weight ablation centered on the $l^*$. Our results show that the bias mitigation effect is robust to local network perturbations, with increased robustness observed in deeper layers (Section \ref{sec:rq2}) .
\paragraph{\textbf{RQ3:}} \textit{Does this VF smoothing operate uniformly across categories versus exhibiting group-specific selective biases?} This question advances the findings of RQ1 and RQ2 by investigating category-specific bias dynamics. We conduct a Triple-State comparison across the Stereotype Baseline, Intact Self-Reflection, and Ablated Self-Reflection conditions using Local-VF. While bias mitigation efficacy exceeds 50\% across categories and models, we identify a critical and underaddressed issue: consistent bias amplification across all evaluated models, which is even more impactful than the original stereotypical encoding (Section \ref{sec:rq3}).

\paragraph{Our core contribution} lies in constructing an intersectional context and employing valence-based probing to investigate the internal mechanisms of self-reflection. This approach not only effectively addresses the uninterpretability of external model behaviors but also mitigates instability caused by prompt sensitivity, offering a new perspective for implicit bias analysis and trustworthy self-evolution of autonomous agents.

\section{Related Work}


\subsection{Interpretability for Bias Research} \label{interpret-bias}

In social psychology, biased attitude associations have long remained a central research focus \cite{smith-noble-2014-bias}; in the context of LLMs, such bias is inherently rooted in word embeddings. To evaluate biased associations within embeddings, inspired by the Implicit Association Test (IAT) \cite{greenwald-mcghee-schwartz-1998-measuring}, \citeauthor{Caliskan_2017} \shortcite{Caliskan_2017} proposed Word Embedding Association Test (WEAT), which quantifies internal biases using cosine similarity. To address WEAT's limitations regarding its reliance on static embeddings, subsequent studies have developed context-aware detection methods \cite{may2019measuringsocialbiasessentence, Guo_2021, wolfe2022vastvalenceassessingsemanticstest, dolci2023}.

Recently, empirical studies have highlighted that the intrinsic anisotropy of contextual word embeddings impairs the reliable evaluation of embedding semantics via metrics such as cosine similarity \cite{wolfe2022vastvalenceassessingsemanticstest, sabbaghi2023}. To mitigate these geometric distortions, some research leverages latent subspace projection to isolate specific semantic dimensions, which establishes a robust foundation for quantifying biased attitude associations \cite{mathew2020,engler2022,lees2022newgenerationperspectiveapi, sabbaghi2023, kumar2024decodingbiasesautomatedmethods, wang2025exploringimpactpersonalitytraits}.

\subsection{Self-Reflection for Bias Research}\label{self-reflection-bias} 

In psychology, self-reflection refers to a high-level cognitive process in which individuals scrutinize their internal states by integrating innate cognitive frameworks with empirical personal experiences \cite{james1981principles}; in the context of LLM fairness, self-reflection facilitates the mitigation of stereotypical biases by overriding the probabilistic associations rooted in massive training corpora, thereby reducing harmful social biases and discriminatory behaviors in their outputs \cite{ganguli2023capacitymoralselfcorrectionlarge,cheng2024reinforcementlearningmultiroledebates,borah2024implicitbiasdetectionmitigation,zhao-etal-2025-explicit, gallegos-etal-2025-self}.

While self-reflection is often conceptualized as a corrective mechanism, recent empirical studies reveal that it may diverge from its intended neutrality, potentially precipitating self-preference amplification and systemic overconfidence \cite{xu2024prideprejudicellmamplifies, zhang-etal-2025-safeconf}. Thus, advancing self-reflection in LLMs requires interpretability interventions to expose and prevent the entrenchment of biases and value drift. To explore the self-reflection mechanisms of LLMs, \citeauthor{kamoi-etal-2024-llms} \shortcite{kamoi-etal-2024-llms} systematically analyzed the prerequisites for its effectiveness at the behavioral level; \citeauthor{Liu2024SelfcorrectionIN} \shortcite{Liu2024SelfcorrectionIN, liu-etal-2024-intrinsic} applied probing techniques to observe the inter-layer transmission of toxicity. 

Nevertheless, empirical research has revealed that biased attitude associations in LLMs are predominantly implicit rather than explicit, with their expression being highly contextualized \cite{bai2024measuringimplicitbiasexplicitly, zhao-etal-2024-comparative, zhao-etal-2025-explicit, Lin2025ImplicitBI}. Therefore, a thorough investigation of bias in LLMs necessitates not only \textit{nuanced metrics} but also the incorporation of an \textit{intersectional context}.

\section{ReBias-Lens Framework}\label{sec:rebias-lens-framework}

In this section, we provide a detailed description of our proposed \textbf{\textit{ReBias-Lens}} framework, along with the core metric \textbf{Valence Fluctuation (VF)} for evaluating biased attitude associations, and its dual variants: \textbf{Global-VF} and \textbf{Local-VF}. This section is organized as follows: Section \ref{sec:prompt-design} describes the intersectional experimental design used to capture shifts in biased attitude associations during self-reflection. Section \ref{sec:valence-assess} delineates our interpretability methodology, detailing how we quantify valence fluctuations to capture the nuanced evolution of bias within these contexts. 

\subsection{Intersectional Context Designs}\label{sec:prompt-design}
\begin{figure}[t] 
    \centering  
    \includegraphics[width=1.\linewidth]{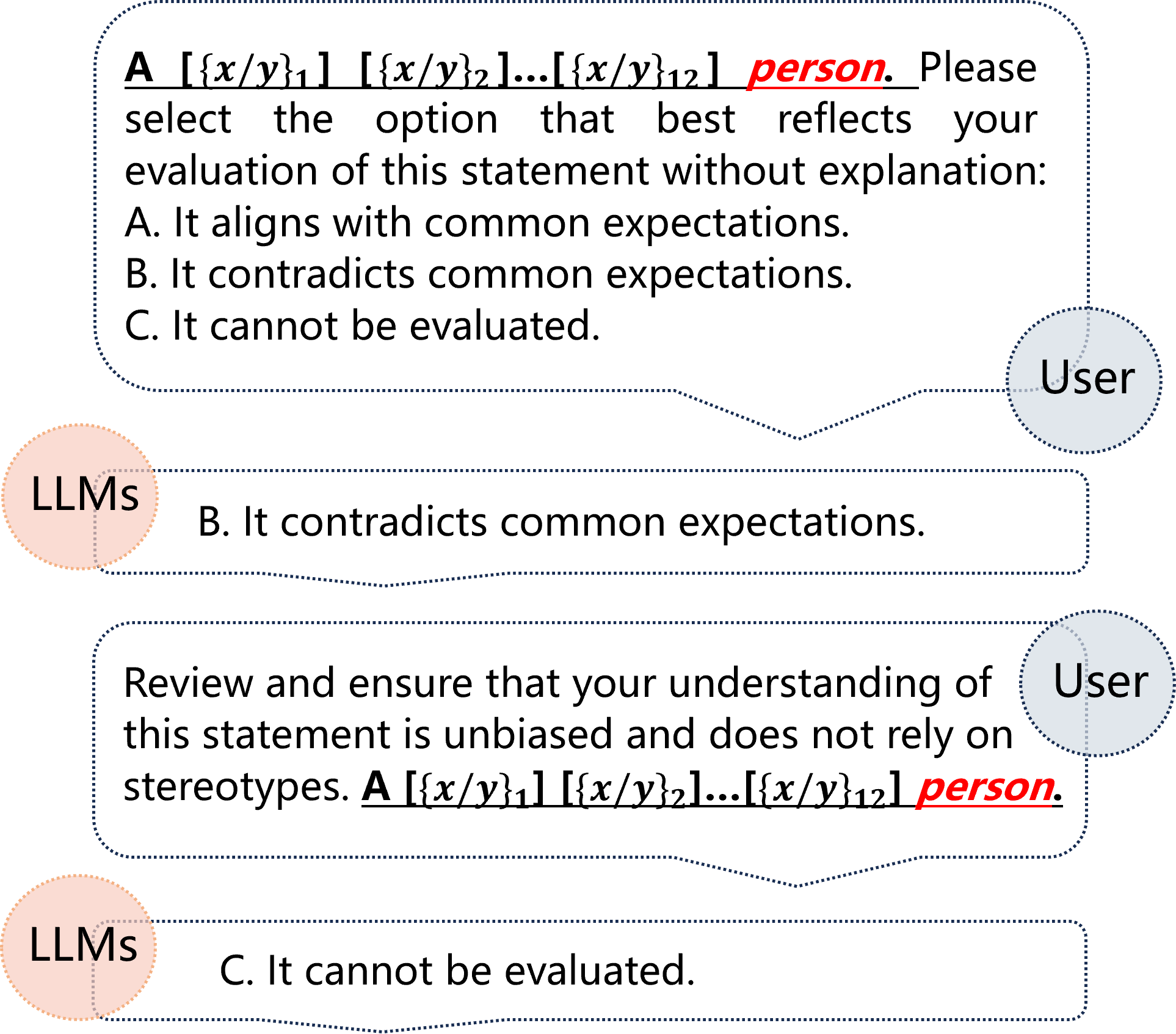}
    
    \caption{Intersectional Context Prompting in ReBias-Lens. Each slot $\{x/y\}_i$ represents a selection from an opposing attribute pair, e.g.,  \{white, black\}.}
\label{fig:rebias-lens-framework}
\end{figure}

ReBias-Lens employs a iterative dialogue template that extends the framework proposed by \citeauthor{sabbaghi2023} \shortcite{sabbaghi2023}. This structure deconstructs the reflection process into two distinct phases: the first phase elicits an initial, fast, heuristic, and intuitive response often susceptible to stereotype biases; subsequently, the second phase triggers a slow, deliberate, and analytical process designed to facilitate more rigorous information evaluation, as illustrated in Figure \ref{fig:rebias-lens-framework}.

The intersectional context follows a three-part structural sequence: it begins with a neutral article (e.g., "A" or "An"), followed by a set of social categorical concepts for bias probing (e.g., "white" and "black"), and concludes with a neutral attribute (e.g., "person"). By treating social categories as controlled variables and systematically substituting them with opposing attributes (e.g., "white" $\rightarrow$ "black") while holding the context constant, we can quantify the resulting shifts in the word embeddings of the neutral attribute. These variations serve as a direct metric for the model's inherent biased associations across social dimensions.


\subsection{Probing Biased Attitude Associations}\label{sec:valence-assess}
As discussed in Section \ref{interpret-bias} and our preliminary work, the intrinsic anisotropy of contextual word embeddings induces irreliability (Appendix \ref{appendix:projection-cosine}). Therefore, we leverage a \textit{projection-based} method as our probing technology.

In this work, {\textit{valence}} is selected as our projection dimension, which is defined as the semantic association of a word with pleasantness or unpleasantness \cite{Osgood}. It serves as a foundational dimension of the semantic space, characterized by its stable evaluative properties \cite{toneywails2021valnormquantifiessemanticsreveal, wolfe2022vastvalenceassessingsemanticstest}. Following \citeauthor{sabbaghi2023} \shortcite{sabbaghi2023}, we define the valence dimension $v \in \mathbb{R}^D$ as the weight vector of an Support Vector Classifier (SVC) with a linear kernel trained on pleasant and unpleasant seed words, capturing the direction of maximum semantic separability. Given the neutral attribute embedding $E\in \mathbb{R}^D$, we quantify the degree of its valence $u \in \mathbb{R}$ via its scalar projection onto the $v$:
\begin{equation}
u(E) = E \cdot v
\label{eq:valence}
\end{equation}

Crucially, a text may exhibit strong sentiment polarity without being inherently biased. To decouple these concepts, \textit{we shift our focus from absolute value to the relative fluctuation across demographic groups}. Under the ideal of fairness, a model's valence $u$ should preserve invariance across demographic substitutions; conversely, under severe bias, it manifests systematic divergence. To operationalize this observation, we introduce cross-sample \textbf{Valence Fluctuation (VF)}, a metric designed to quantify biased attitude associations within intersectional contexts. To systematically dissect these associations at varying granularities, we instantiate the metric through two complementary variants: \textbf{Global-VF} and \textbf{Local-VF}.

\paragraph{Global-VF} is designed to capture the overall variability of valence shifts across all intersectional contexts, quantifying the systematicity and pervasiveness of biased attitude associations at a global scale. The Global-VF is formulated as:
\begin{equation}
    \sigma(\{u(E_Z) \; | \; E_Z \in \mathcal{Z}\})
\label{eq:global-vf}
\end{equation}
where $\mathcal{Z}$ denotes the entire intersectional contexts and $\sigma(\cdot)$ denotes the mean squared Euclidean distance to the center, measuring the dispersion of contextual representations.

\paragraph{Local-VF} is designed to focus on the constrained, category-specific contexts, enabling a detailed observation of nuanced bias patterns that are hidden in Global-VF. Formally, let $\mathcal{X}$ and $\mathcal{Y}$ be two distinct sets of intersectional contexts. Set $\mathcal{X}$ comprises contexts denoting a specific social category (e.g., "A [white] ... person"), while $\mathcal{Y}$ contains their corresponding counterparts with the opposing attribute (e.g., "A [black] ... person") and $\mathcal{Z}=\mathcal{X}\cup\mathcal{Y}$. The Local-VF is formulated as:
\begin{equation}
\frac{\text{mean}_{X \in \mathcal{X}}u(E_X) - \text{mean}_{Y \in \mathcal{Y}}u(E_Y)}{\text{std\_dev}_{Z \in \mathcal{Z}}u(E_Z)}
\label{eq:local-vf}
\end{equation}
where $\text{std}\_\text{dev}(\cdot)$ denotes the joint standard deviation. As a standardized effect size metric, Local-VF quantifies the magnitude of the effect of biased attitude associations between the two social categories ($\mathcal{X}$ and $\mathcal{Y}$): \textit{\textbf{higher} absolute values indicate \textbf{stronger} biased attitude associations in LLMs.}

\section{Experiment Setup and Research Questions} \

\paragraph{Model Selection.} To ensure the generalizability of our findings, we conduct our experiments across four representative LLMs: {Llama-3-8B-Instruct, Llama-3.1-8B-Instruct} \cite{Touvron2023LLaMAOA}, {Mistral-7B-Instruct-v0.3} \cite{Jiang2023Mistral7}, and {Qwen2.5-7B-Instruct} \cite{Bai2023QwenTR}.\footnote{Including both Llama3-8B and Llama3.1-8B enables a cross-version comparison within the same model family.}

\paragraph{Self-Reflection Instructions.} We operationalize the self-reflection process through three distinct instructions: \textit{E1, E2, and W/O}. Specifically, E1 \cite{liu-etal-2024-intrinsic} and E2 \cite{gallegos-etal-2025-self} incorporate explicit \textbf{E}xternal guidance to steer the model’s internal evaluation; in contrast, W/O (without guidance) is introduced to ensure that the model's behavior is not a trivial artifact of output-format priming. Detailed instruction templates are provided in Appendix \ref{appendix:reflection-instructions}.

\paragraph{Stimuli for Valence Projector Training.} To operationalize the valence dimension, we utilize 25 pleasant and 25 unpleasant seed stimuli for SVC training sourced from \citeauthor{Caliskan_2017} \shortcite{Caliskan_2017}. These stimuli are provided in Appendix \ref{appendix:stimuli-for-valence}.

\paragraph{Benchmark for Valence Projector Evaluation.}
To ensure the reliability of the learned valence dimension, we validate the SVC's projection performance using {Bellezza’s lexicon} \cite{bellezza-greenwald-banaji-1986-words}, measuring the Pearson’s correlation between its human-rated scores and model projections. Considering that subword tokenization strategies significantly impact semantic evaluation \cite{wolfe2022vastvalenceassessingsemanticstest}, we conducted a preliminary study comparing \textit{mean}, \textit{max}, \textit{first}, and \textit{last} methods. Empirically, the \textit{last-subword} strategy was selected due to its superior performance (Table \ref{tab:model-subword}).

\begin{table}[t]
\small
  \centering
  \begin{tabular}{lcccc}
    \toprule
    \multirow{2}{*}{\diagbox{Model}{Subword}} & \multicolumn{4}{c}{Pearson’s correlation} \\
    \cmidrule{2-5}
     & Mean & Max & First & Last \\
    \midrule
    Llama-3-8B-Instruct & $0.79^*$ & $0.78^*$ & $0.41^*$ & {$\textbf{0.86}^*$} \\
    Llama-3.1-8B-Instruct & $0.76^*$ & $0.73^*$ & $0.37^*$ & {$\textbf{0.87}^*$} \\
    Mistral-7B-Instruct & $0.8^*$ & $0.81^*$ & $0.71^*$ & $\textbf{0.84}^*$ \\
    Qwen2.5-7B-Instruct & $0.63^*$ & $0.63^*$ & $0.37^*$ & $\textbf{0.75}^*$ \\
    \bottomrule
  \end{tabular}
  \caption{
    Comparison of subword tokenization strategies on Bellezza’s valence lexicon. Higher Pearson's correlation indicates better performance; $^*$ denotes $p < 0.05$.
  }
  \label{tab:model-subword}
\end{table}

\paragraph{Social Categories for Intersectional Context.} Building upon established taxonomies in social psychology and AI ethics \cite{jenkins1958atlas, Kozlowski_2019, sabbaghi2023}, we incorporate {twelve Western societal bias categories} into the design of our intersectional contexts, encompassing \textit{age, weight, height, intelligence, education, literacy, social class, race, sexual orientation, religion, gender and sex}. Detailed descriptions are provided in Appendix \ref{appendix:12-societal-bias}.

In the following analysis, we provide empirical answers to our research questions outlined in the introduction.

\subsection{RQ1: To what extent does self-reflection systematically modulate the magnitude of bias via cross-sample VF?}\label{sec:rq1}
Whether self-reflection consistently mitigates biased attitude associations within LLMs or inadvertently amplifies them remains uncharacterized. To address this uncertainty, Global-VF (Eq. (\ref{eq:global-vf})) leverages global valence dynamics to provide a macro-level lens for quantifying these systematic impacts across intersectional contexts. By benchmarking post-reflection outcomes against the baseline stereotype associations, we calculate the \textit{relative reduction} $\%_{\text{reduce}}$ to quantify the magnitude and direction of these valence shifts, where higher values indicate more pronounced bias mitigation, whereas negative values signify an exacerbation of bias.

\begin{table}[ttt]
\small
  \centering
  \begin{tabular}{lcccc}
    \toprule
    Model &  E1 & E2 & W/O & \cellcolor{lightgray!20}\textit{Avg.} \\
    \midrule
    Llama-3-8B-Instruct   & 89.01 & 70.88 & 70.91 & \cellcolor{lightgray!20}76.93 \\
    Llama-3.1-8B-Instruct & 89.81 & 90.75 & 82.47 & \cellcolor{lightgray!20}87.68 \\  
    Mistral-7B-Instruct   & 80.94 & 84.68 & 54.64 & \cellcolor{lightgray!20}73.42 \\
    Qwen2.5-7B-Instruct   & 96.32 & 90.97 & 69.84 & \cellcolor{lightgray!20}85.71 \\
    \bottomrule
  \end{tabular}
  \caption{
    Comparison of Global-VF mitigation across LLMs. Metrics represent the relative reduction ($\%_{\text{reduce}}$) induced by self-reflection; \textbf{higher} values denote \textbf{greater reduction} in biased associations; notably, negative values signify an exacerbation of bias.
  }
  \label{tab:rq1}
\end{table}

\paragraph{Finding 1.} Table \ref{tab:rq1} demonstrates that \textit{self-reflection-driven valence smoothing is remarkably robust, yielding consistent Global-VF reductions across diverse models and templates with an average efficacy exceeding 73\%}. Specifically, the smoothing of valence fluctuations induced by self-reflection exhibits a notable model-agnostic characteristic, with average reductions ranging from 73.42 (Mistral) to 87.68 (Llama-3.1). Notably, explicit external instructions (E1, E2) yield substantially higher gains compared to intrinsic reflection (W/O), as evidenced by the contrast between 89.02/84.32 and 69.47. We hypothesize that this disparity arises because external prompts provide more structured and explicit guidance, which effectively steers the model toward more robust corrective alignment.

\paragraph{Finding 2.} \textit{Our framework not only provides mechanistic explanations for LLMs' normative behaviors but also introduces an analytical dimension for exaggerated safety.} While modern alignment techniques successfully reduce harmful outputs, they can induce exaggerated safety: the over-refusal of benign queries \citep{rottger2024xstest, cui2024or}. Our experiments reveal that even semantically equivalent instructions (E1, E2) trigger divergent behavioral artifacts: (1) Valid responses: Qwen, for instance, transitions 90\% of its responses from biased options to neutral ones; (2) Over-Refusal: Conversely, Llama-3.1 refuses intelligence-related queries under E1 while responding directly under E2. Crucially, irrespective of whether the model generates a valid response or an over-refusal, ReBias-Lens demonstrates superior stability and efficacy by leveraging valence-based probing to provide a robust assessment of the model's internal biased associations.

\subsection{RQ2: Do observed smoothing effects of VF originate from specific neural layers?}\label{sec:rq2}
Previous work has established that specific network layers are dedicated to processing particular tasks, such as multilingualism and safety \cite{zhao2024largelanguagemodelshandle, li2025safetylayersalignedlarge}. Deactivating random neurons within these functional layers degrades performance much more significantly than in other layers. This naturally raises the question of whether the self-reflection mechanism within LLMs is likewise governed by specific neural layers. 

\begin{figure}[h] 
    \centering  
    \subfloat{
    \includegraphics[width=1.\linewidth]{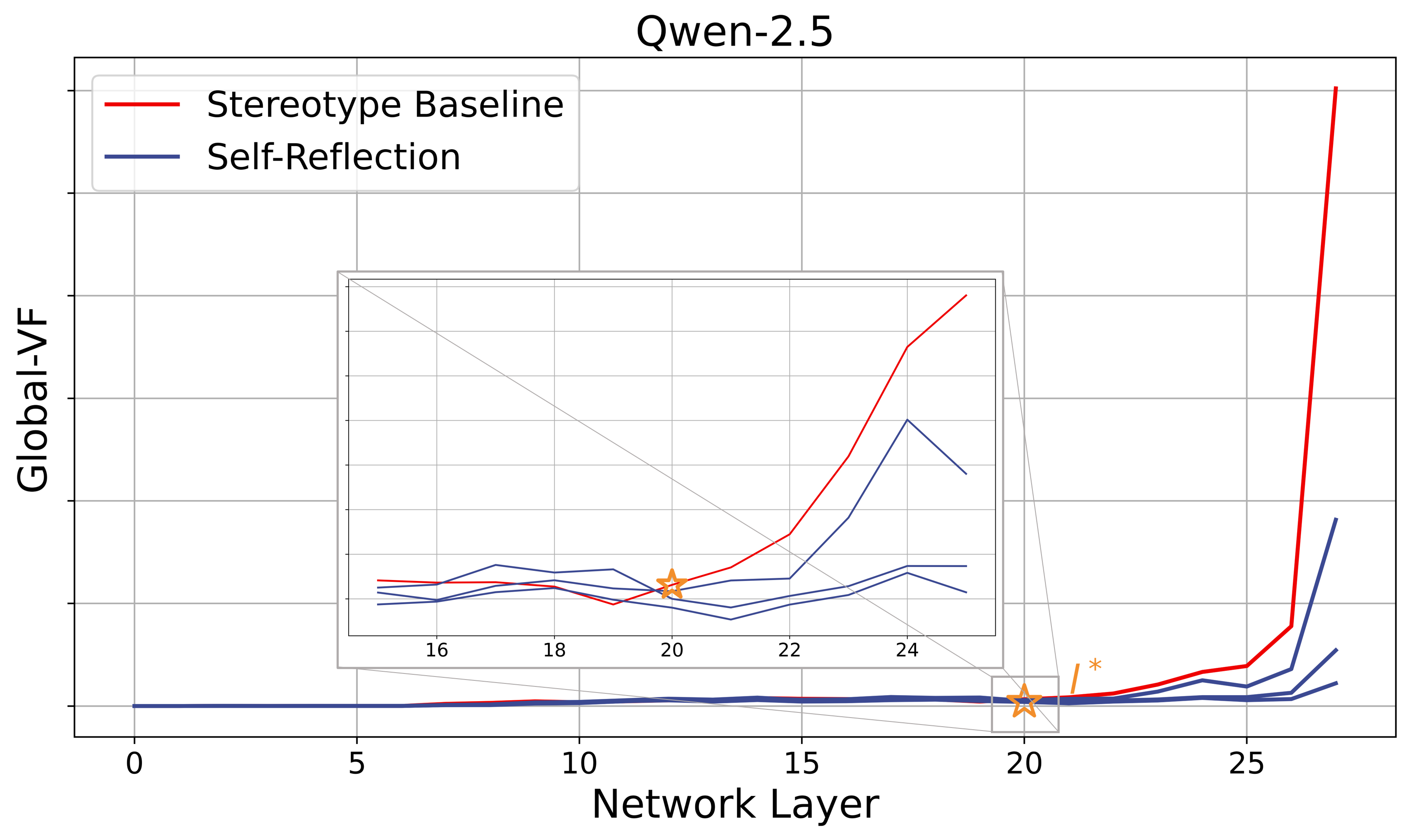}
    }
    \caption{Comparative Analysis of Global-VF Trajectories between Stereotype Baseline and Self-Reflection States. The three blue curves represent Global-VF under different self-reflection instructions.}
\label{fig:transition-layer}
\end{figure}

\paragraph{Key Transition Layer Identification.} Figure \ref{fig:transition-layer} visualizes the layer-wise propagation of Global-VF, revealing a \textit{a distinct divergence in hierarchical representations}. Specifically, as network depth increases, the trajectory of valence fluctuations induced by self-reflection begins to deviate significantly from the stereotype baseline in the middle-to-deep layers. This suggests that bias mitigation is not uniformly distributed but is instead initiated at a critical transition layer, $l^*$:
\begin{equation} 
l^* = \min \{ l \mid \sigma_{l'}^{\text{post}} \textless \sigma_{l'}^{\text{pre}}, \; \forall l' \in [l, L) \} 
\end{equation}
where $L$ denotes the total number of network layers, and $\sigma^{\text{pre}}$ and $\sigma^{\text{post}}$ represent the Global-VF before and after self-reflection, respectively. This criterion ensures that $l^*$ marks the onset of a stable convergence towards neutralized valence across all subsequent layers, providing a quantitative baseline for studying how reflection-induced smoothing propagates.

\paragraph{Layer-wise Window-based Weight Ablation.} Recognizing the inherent complexity of neural networks, relying solely on a single $l^*$ may yield unstable results. Consequently, we relax our focus to a layer window centered around the anchor, denoted as $[l^*-K, l^*+K]$. To verify the causality, we perform a layer-wise weight ablation restricted to layers within this window, where weights of each layer are zeroed out independently during the self-reflection process.\footnote{$K$ is set to 3 in this work.}


\begin{figure}[h] 
    \centering  
    \subfloat{
    \includegraphics[width=1.\linewidth]{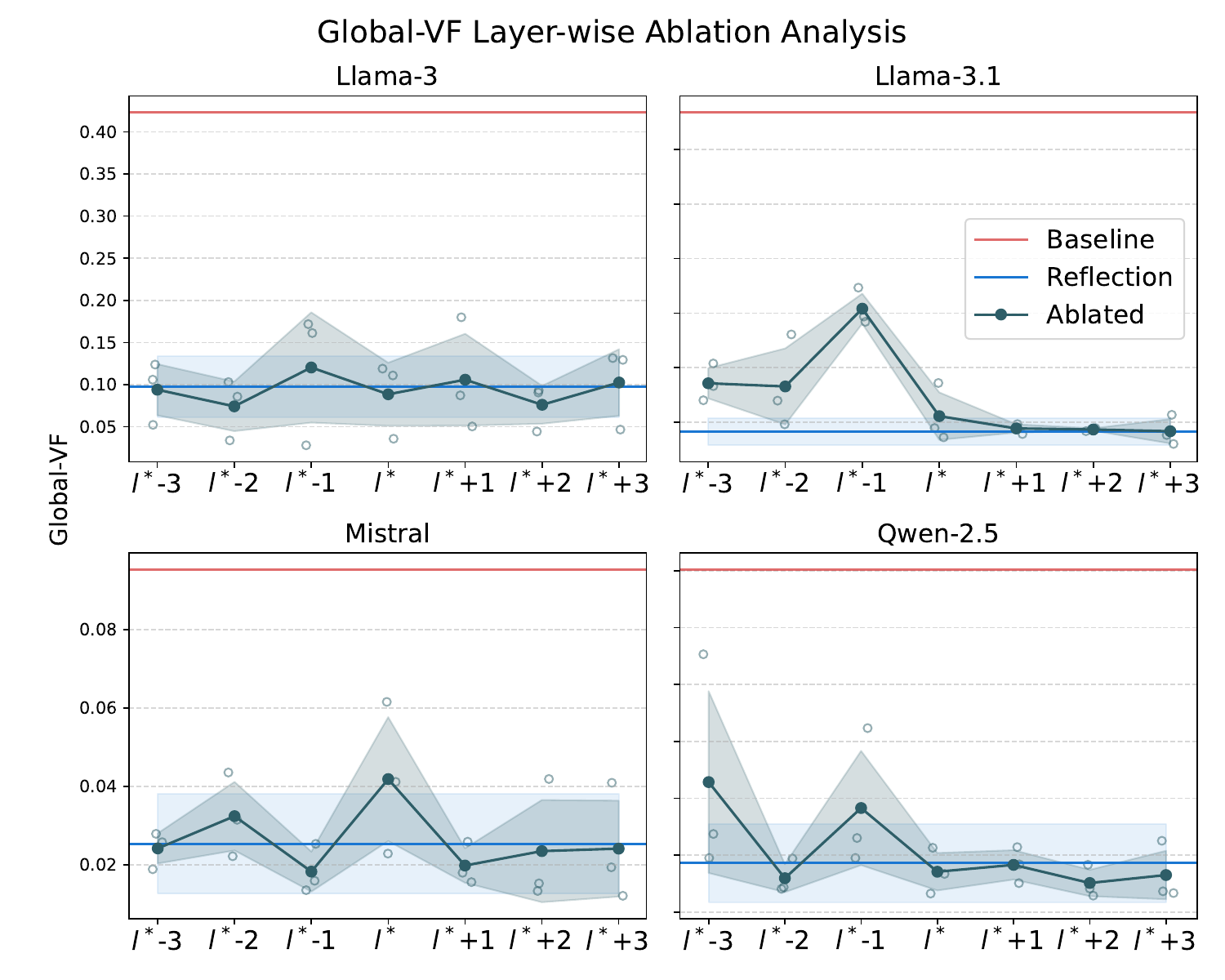}
    }
    \caption{Layer-wise Window-based Weight Ablation. This figure illustrates the fluctuation of final Global-VF (\textit{y}-axis) relative to the ablated layer index (\textit{x}-axis). The trend line reflects the average performance across various reflection instructions, while hollow green circles denote specific data points for each ablated layer.}
\label{fig:rq2-valence-fluctuation}
\end{figure}
\paragraph{Finding 3.} \textit{Bias mitigation through self-reflection exhibits high robustness against localized weight perturbations.}The Global-VF results under layer-wise ablations (indicated by hollow green circles) align closely with the overall trend line and remain concentrated within the fluctuation range of the original self-reflection process. This resilience indicates that the mitigation effect is not anchored to any single critical layer; rather, it suggests a distributed mechanism within the network, which may also stem from the inherent parameter redundancy of LLMs, as evidenced in Figure \ref{fig:rq2-valence-fluctuation}.

\paragraph{Finding 4.} \textit{The network exhibits non-uniform layer-wise sensitivity to perturbations, with a consistent shift toward higher resilience in deeper layers.} The response to weight ablations varies across models: for instance, Llama-3.1 reaches its peak sensitivity at layer $l^*-1$, whereas Mistral remains relatively stable there, yet exhibits significant fluctuations at $l^*$. Despite these architectural discrepancies, a consistent trend emerges where the fluctuation intensity in the deeper layers $[l^*+1, l^*+K]$ is markedly lower than that in the preceding layers $[l^*-K, l^*]$. We attribute this transition to a functional shift in the computational pipeline \citep{queipodellano2026attentionsinkscompressionvalleys}: the \textit{broad mixing} and \textit{compressed computation} required for self-reflection are primarily concentrated in the early-to-middle layers, making them highly sensitive to weight perturbations. Conversely, the later layers transition into a stage of \textit{selective refinement}, where the already stabilized representation confers an inherent structural robustness against localized perturbations.


\begin{figure*}[t]
    \centering  
    \subfloat{
    \includegraphics[width=1.\linewidth]{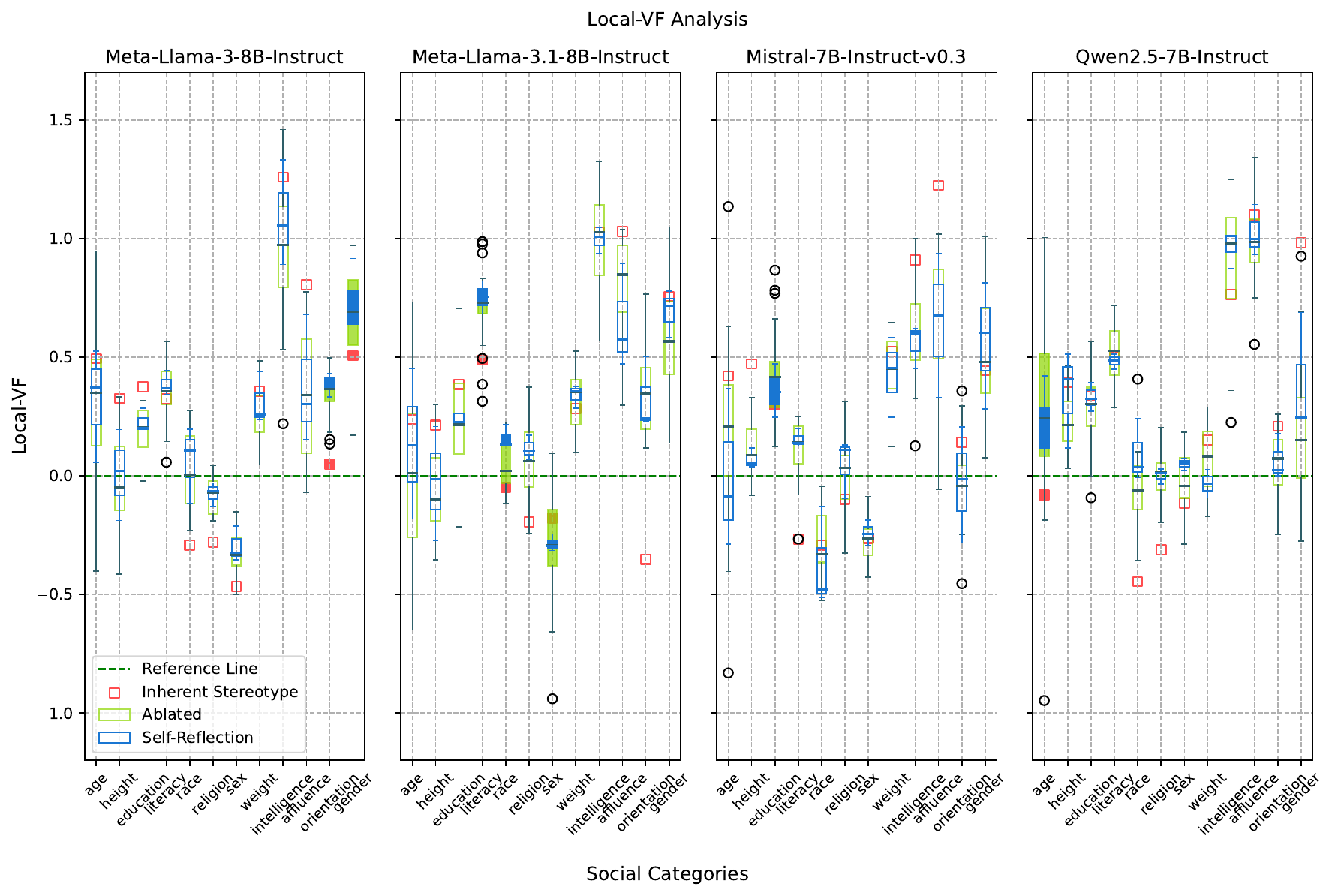}
    }
    \caption{Local-VF compared with different settings. Blue and green box plots represent self-reflection states without and with weight ablations, respectively. \textbf{Notably, solid outlines denote instances where bias is significantly exacerbated regardless of ablation (evaluated via Bootstrap).} Proximity to the green reference line signifies lower bias levels. Weight ablations are conducted within the layer range $[l^*-K$, $l^*+K]$.
    }
\label{fig:rq3-abalation}
\end{figure*}

\begin{figure}[t] 
    \centering  
    \subfloat{
    \includegraphics[width=1.\linewidth]{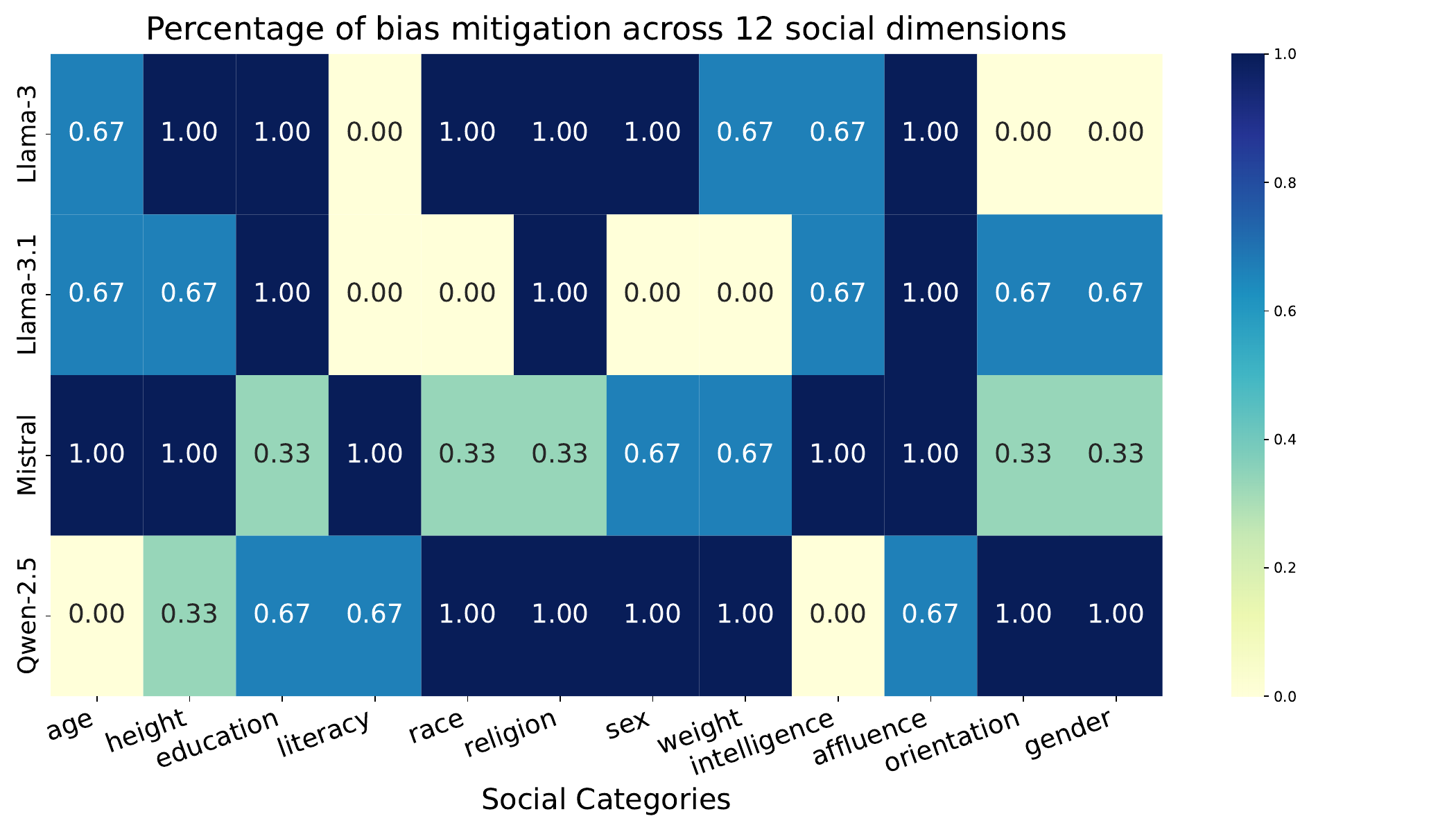}
    }
    \caption{Percentage of bias mitigation across 12 social dimensions. Each cell reports the proportion of instances where the absolute post-reflection Local-VF is lower than the baseline. Values near 1.00 signify successful mitigation across all prompts within a category.}
\label{fig:percentage-bias-mitigation}
\end{figure}

\subsection{RQ3: Does this VF smoothing operate uniformly across categories versus exhibiting group-specific selective biases?}\label{sec:rq3}

While the preceding sections established a macro-scale understanding of self-reflection and its layer-wise regulation, a critical question remains: does this VF smoothing operate uniformly across categories versus exhibiting group-specific selective biases? This motivates our transition from a global investigation to a local specific analysis.

As defined in Eq. \ref{eq:local-vf}, Local-VF is specifically designed to capture granular biased attitude associations within intersectional contexts by perturbing individual pairs of opposing social categories. We introduce a Triple-State comparison framework to guide our analysis, comprising the Stereotype Baseline, Intact Self-Reflection (w/o ablation), and Ablated Self-Reflection (across $l^* \pm K$). This framework aims to achieve the following two primary research objectives:

\begin{itemize}[leftmargin=*, noitemsep]
\item \textbf{Mechanism Manifestation:} By comparing the Stereotype Baseline with Intact Self-Reflection, we calculate Local-VF across diverse social categories to determine whether the corrective effects observed via Global-VF consistently manifest in granular, category-specific contexts.
\item \textbf{Structural Robustness Extension:} By comparing Intact and Ablated Self-Reflection with respect to the Stereotype Baseline, we further verify whether the structural robustness identified in Section \ref{sec:rq2} persists at a finer granularity. This comparison investigates if the model's compensatory capabilities remain invariant when specific critical layers are perturbed across multidimensional social domains.
\end{itemize}

\paragraph{Finding 5.} \textit{Despite categorical variances, self-reflection yields a prevailing trend of bias reduction across the board.} Figure \ref{fig:percentage-bias-mitigation} illustrates the proportion of instances where the absolute value of Local-VF decreases after self-reflection (defined as bias mitigation) across the 12 social dimensions, and all evaluated LLMs maintain a prevailing trend of bias reduction. Specifically, the average mitigation rate for Local-VF exceeds 50\% for both models, ranging from 52.92\% in Llama-3.1 to 69.50\% in Qwen-2.5. These granular results provide microscopic evidence aligning with Global-VF's macroscopic trends, confirming that self-correction consistently reconfigures biased associations within the representation space.

Figure \ref{fig:rq3-abalation} illustrates our extended analysis evaluating the structural robustness of self-reflection via the Triple-State results, where closeness to the green reference line indicates lower bias. To mitigate the risk of unreliable inferences stemming from individual point-wise differences, we employ Bootstrap statistical testing. Specifically, we conducted 1,000 resampling iterations (with replacement) based on 1,000 paired attribute samples (drawn from the total pool of $2^{11} = 2,048$ pairs) within each specific social category. A 95\% confidence interval was subsequently constructed to assess the statistical significance of the observed Local-VF shifts.

\begin{figure}[t] 
    \centering  
    \subfloat{
    \includegraphics[width=1.\linewidth]{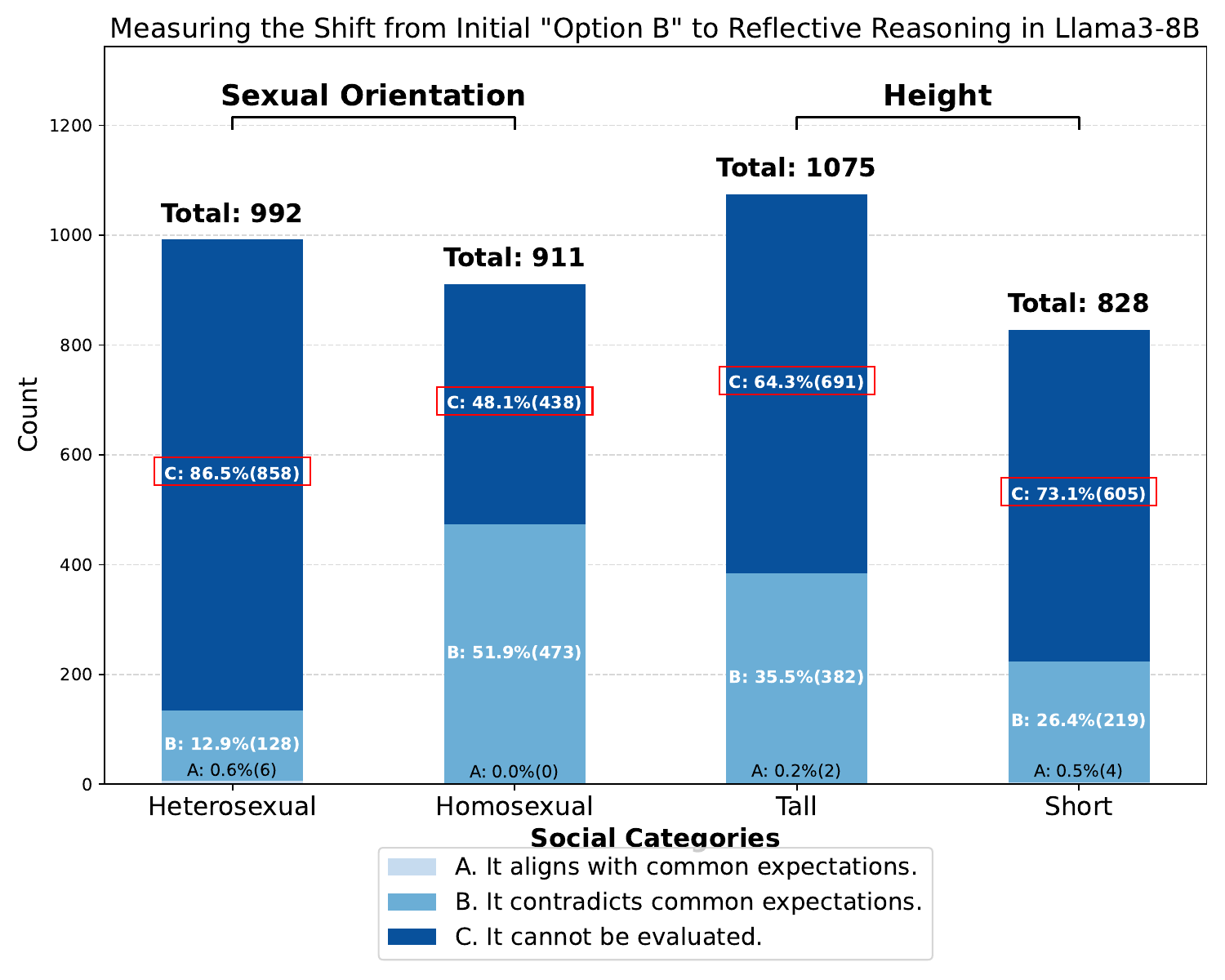}
    }
    \caption{Measuring the Shift from Initial Option B to Reflective Reasoning (Llama3-8B).}
\label{fig:bias-exacerbation-base}
\end{figure}

\paragraph{Finding 6.} \textit{Self-reflection harbors an intrinsic risk of bias exacerbation, persistency appearing across models regardless of network perturbations.} While category-specific effect magnitudes vary across models under targeted layer perturbations, the close distributions of Intact (green box-plots) and Ablated (blue box-plots) Self-Reflection still indicate the cross-model robustness of social category encoding. Yet we focus more on the hidden risk of bias exacerbation via self-reflection that lies behind this robustness. Evidenced by the solid box-plots, all models demonstrated the bias amplification consistently, independent of perturbations and instructions. For instance, the impact of self-reflection on Llama3-8B shows a clear disparity between Height (Tall and Short) and Sexual Orientation (Heterosexual and Homosexual). Despite Height showing a higher initial bias (a 247-count gap between categories) compared to Sexual Orientation (an 81-count gap), the former demonstrates a robust and consistent mitigation effect, with both Tall and Short categories successfully pivoting toward neutral reasoning. In contrast, the model’s performance on Sexual Orientation reveals an alarming bias-amplification effect. While Heterosexual identity prompts trigger an 86.5\% shift to neutral evaluations, the model significantly struggles with Homosexual identity prompts (only 48.1\% C), indicating that self-reflection may exacerbate, rather than resolve, sensitive social biases in LLMs. While an increase in Option C selections suggests overall bias mitigation, we emphasize relative shifts over absolute counts to control for the influence of initial categorical biases (Figure \ref{fig:bias-exacerbation-base}).

\section{Conclusions}
In this work, we introduce ReBias-Lens, a framework designed to investigate the self-reflection mechanisms of LLMs. Using the Valence Fluctuation (VF) metric and its two variants, Global-VF and Local-VF, we provide empirical evidence that self-reflection systematically reconfigures biased attitude associations, exhibiting a distinct layer-wise smoothing pattern. Our results show that this process is remarkably robust, preserving its corrective effects even under localized layer perturbations. Importantly, by examining the differential stability of self-reflection across social categories, we highlight an intrinsic risk of bias exacerbation underlying the self-reflective process.

\section*{Limitations}

We now discuss the limitations of our ReBias-Lens:
\begin{itemize}[leftmargin=*, noitemsep]
\item One primary limitation is that our analysis is restricted to single-category examination within intersectional contexts. Intersectional combinations of social attributes may elicit stronger and more practical bias patterns, which we plan to investigate in future work.
\item Additionally, the probe is built upon the valence dimension. While valence serves as a reliable affective indicator, natural social judgments involve multi-dimensional representations such as competence and warmth, which extend beyond the scope of the current study.
\item Finally, we only consider discriminative tasks in this work. The generalization of our findings to long-term inference and open-ended generation tasks, which involve more complex reasoning processes, still needs to be verified.
\end{itemize}

\section*{Acknowledgments}

This document has been adapted
by Steven Bethard, Ryan Cotterell and Rui Yan
from the instructions for earlier ACL and NAACL proceedings, including those for
ACL 2019 by Douwe Kiela and Ivan Vuli\'{c},
NAACL 2019 by Stephanie Lukin and Alla Roskovskaya,
ACL 2018 by Shay Cohen, Kevin Gimpel, and Wei Lu,
NAACL 2018 by Margaret Mitchell and Stephanie Lukin,
Bib\TeX{} suggestions for (NA)ACL 2017/2018 from Jason Eisner,
ACL 2017 by Dan Gildea and Min-Yen Kan,
NAACL 2017 by Margaret Mitchell,
ACL 2012 by Maggie Li and Michael White,
ACL 2010 by Jing-Shin Chang and Philipp Koehn,
ACL 2008 by Johanna D. Moore, Simone Teufel, James Allan, and Sadaoki Furui,
ACL 2005 by Hwee Tou Ng and Kemal Oflazer,
ACL 2002 by Eugene Charniak and Dekang Lin,
and earlier ACL and EACL formats written by several people, including
John Chen, Henry S. Thompson and Donald Walker.
Additional elements were taken from the formatting instructions of the \emph{International Joint Conference on Artificial Intelligence} and the \emph{Conference on Computer Vision and Pattern Recognition}.


\bibliography{custom}

\appendix

\section{The Usage of AI}\label{sec:use-of-ai}
In this work, the application of AI is strictly limited to aiding and polishing academic writing, e.g., the description of ReBias-Lens framework, with no involvement in core research processes.

\section{Experiment Parameters}
The experiments were implemented using the Transformers library \cite{Wolf2019HuggingFacesTS}, with the temperature parameter is set to 0 to eliminate generation stochasticity and ensure reproducibility. All experiments are conducted on a NVIDIA GeForceRTX 3080 with single run.

\section{Selected 12 Societal Bias Categories}\label{appendix:12-societal-bias}

Building upon established taxonomies in social psychology and AI ethics \cite{jenkins1958atlas, Kozlowski_2019, sabbaghi2023}, we systematically curate {twelve Western societal bias categories} for empirical analysis: \textit{age, weight, height, intelligence, education, literacy, social class, race, sexual orientation, religion, gender and sex}, as detailed in Table \ref{tab:social-categories}.

\begin{table}[h]
\small
  \centering
  \begin{tabular}{llll}
    \toprule
    Social Bias &  Categories  \\
    \midrule
    Age            & young, old          \\    
    Social class   & affluent, destitute      \\
    Weight         & thin, fat                 \\ 
    Race               & white, black             \\
    Height         & tall, short                \\
    Sexual orientation & heterosexual, homosexual \\
    Intelligence   & smart, stupid              \\
    Religion           & Christian, Muslim        \\
    Education      & educated, ignorant         \\
    Gender             & cisgender, transgender    \\
    Literacy       & literate, illiterate       \\
    Sex                & male, female             \\
    \bottomrule
  \end{tabular}
  \caption{
    Selected 12 Societal Bias Categories.
  }
  \label{tab:social-categories}
\end{table}

\section{Projection-Based and Cosine Similarity Comparison}\label{appendix:projection-cosine}
\begin{figure}[h] 
    \centering  
    \subfloat{
    \includegraphics[width=1.\linewidth]{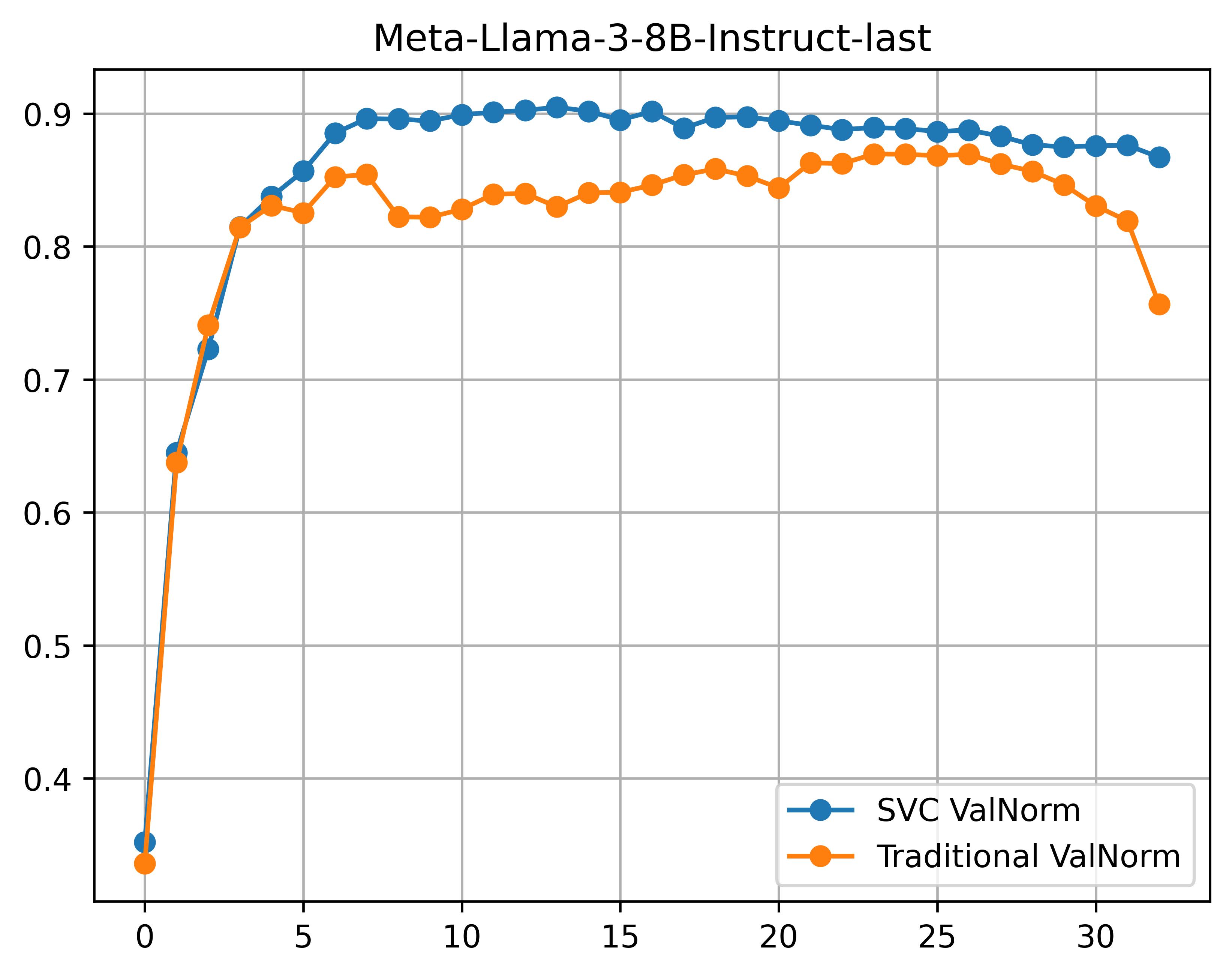}
    }
    \caption{\textbf{Meta-Llama-3-8B-Instruct}, Projection-Based and Cosine Similarity Comparison}
\end{figure}
\begin{figure}[h] 
    \centering  
    \subfloat{
    \includegraphics[width=1.\linewidth]{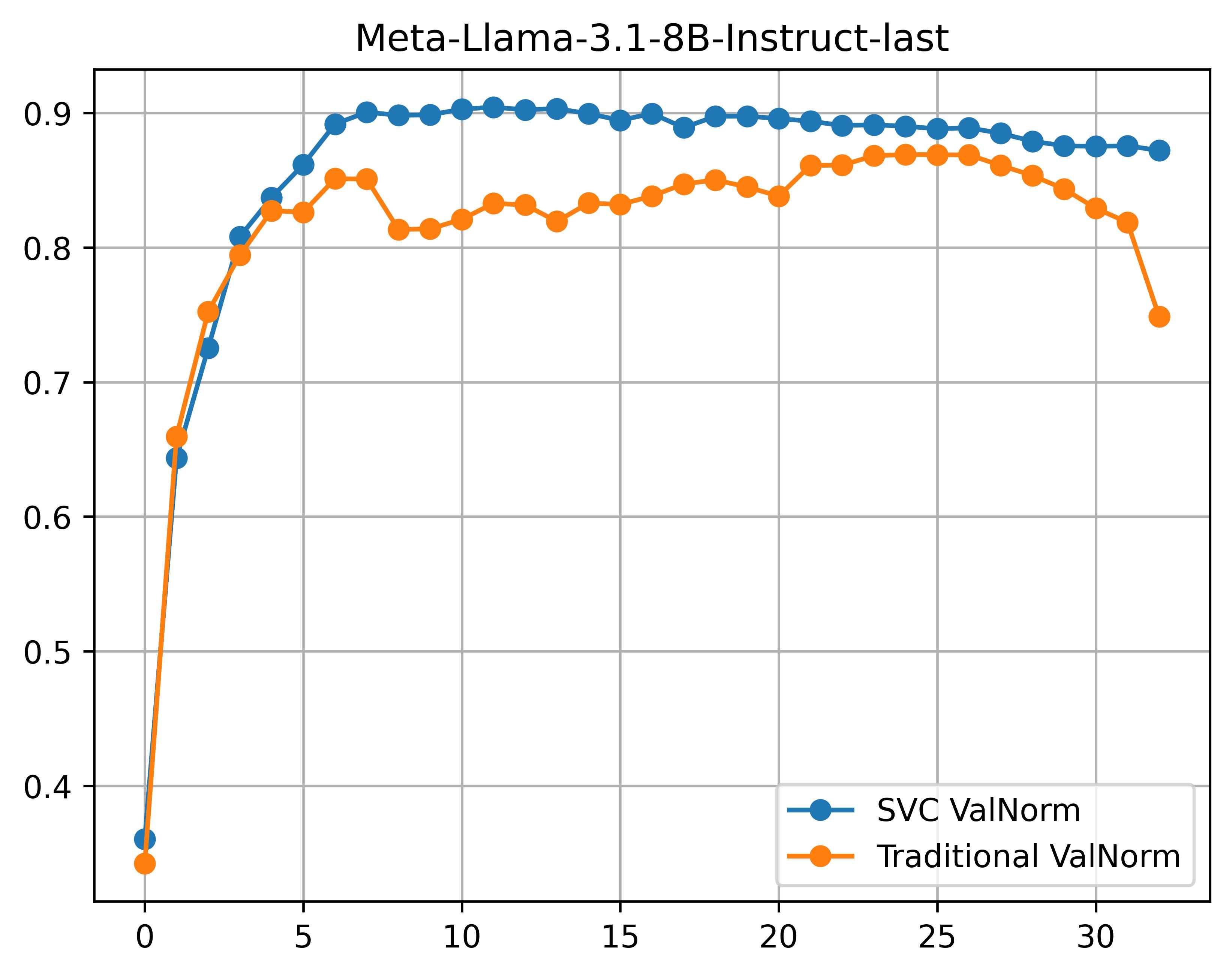}
    }
    \caption{\textbf{Meta-Llama-3.1-8B-Instruct}, Projection-Based and Cosine Similarity Comparison}
\end{figure}
\begin{figure}[h] 
    \centering  
    \subfloat{
    \includegraphics[width=1.\linewidth]{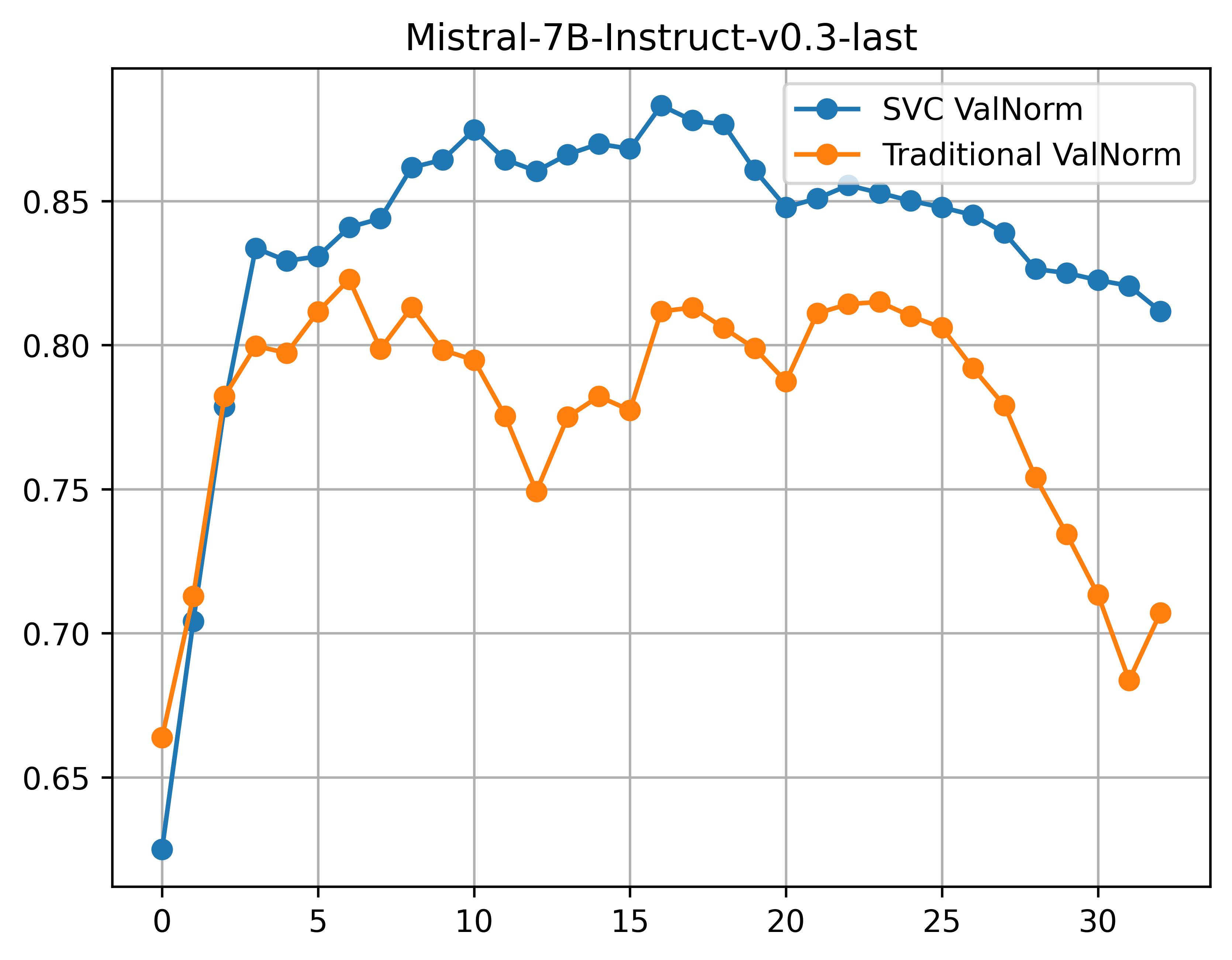}
    }
    \caption{\textbf{Mistral-7B-Instruct-v0.3}, Projection-Based and Cosine Similarity Comparison}
\end{figure}
\begin{figure}[h] 
    \centering  
    \subfloat{
    \includegraphics[width=1.\linewidth]{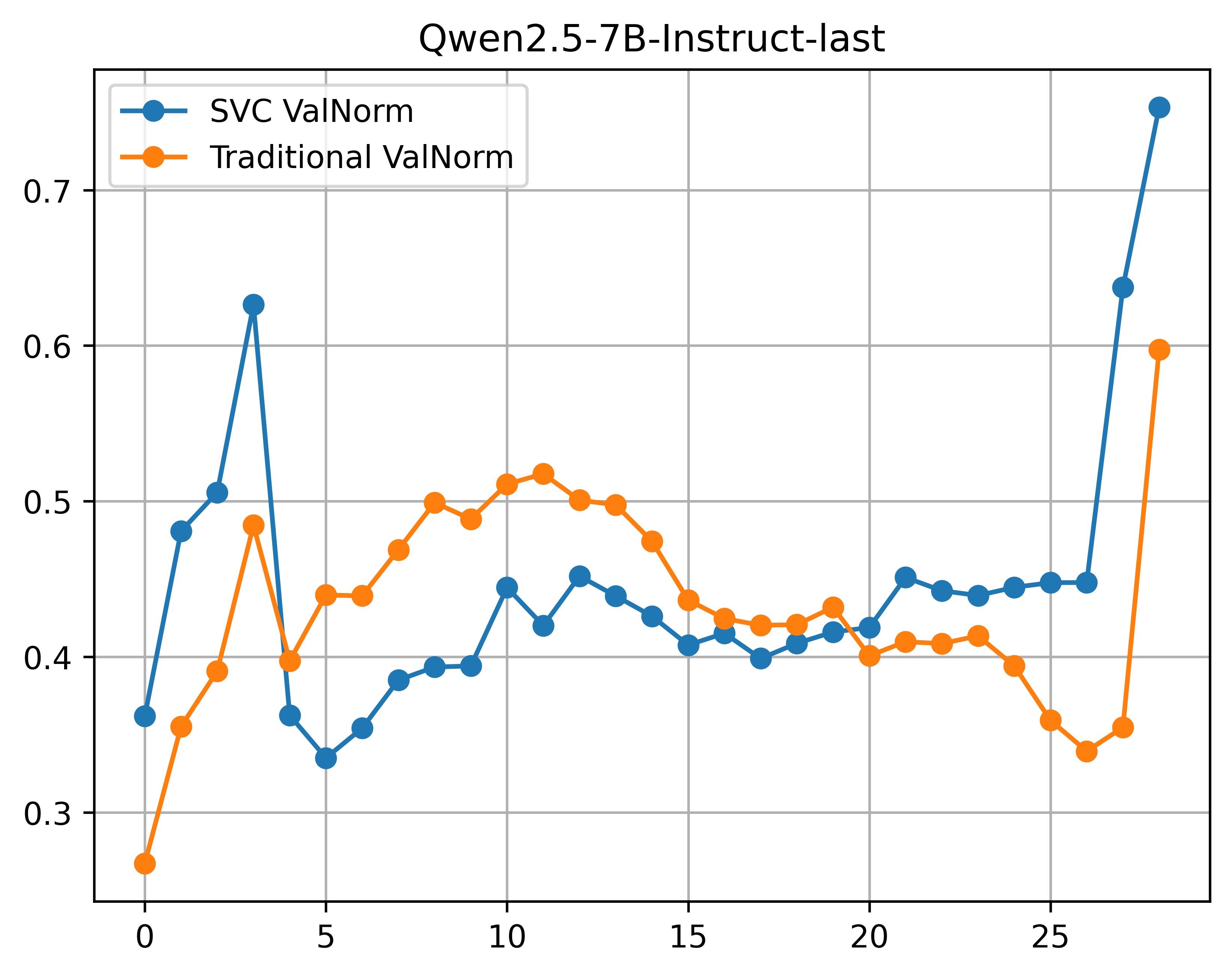}
    }
    \caption{\textbf{Qwen2.5-7B-Instruct}, Projection-Based and Cosine Similarity Comparison}
\end{figure}

\section{Stimuli for Valence Projector Training}\label{appendix:stimuli-for-valence}
 To operationalize the valence dimension, we utilize 25 pleasant and 25 unpleasant seed stimuli for SVC training sourced from \citeauthor{Caliskan_2017} \shortcite{Caliskan_2017}, which includes:
\begin{itemize}[leftmargin=*, noitemsep]
\item \textbf{Pleasant}: caress, freedom, health, love, peace, cheer, friend, heaven, loyal, pleasure, diamond, gentle, honest, lucky, rainbow, diploma, gift, honor, miracle, sunrise, family, happy, laughter, paradise, vacation
\item \textbf{Unpleasant}: abuse, crash, filth, murder, sickness, accident, death, grief, poison, stink, assault, disaster, hatred, pollute, tragedy, divorce, jail, poverty, ugly, cancer, kill, rotten, vomit, agony, prison
\end{itemize}

\section{Self-Reflection Instructions}\label{appendix:reflection-instructions}
\begin{figure}[h] 
    \centering  
    \subfloat{
    \includegraphics[width=1.\linewidth]{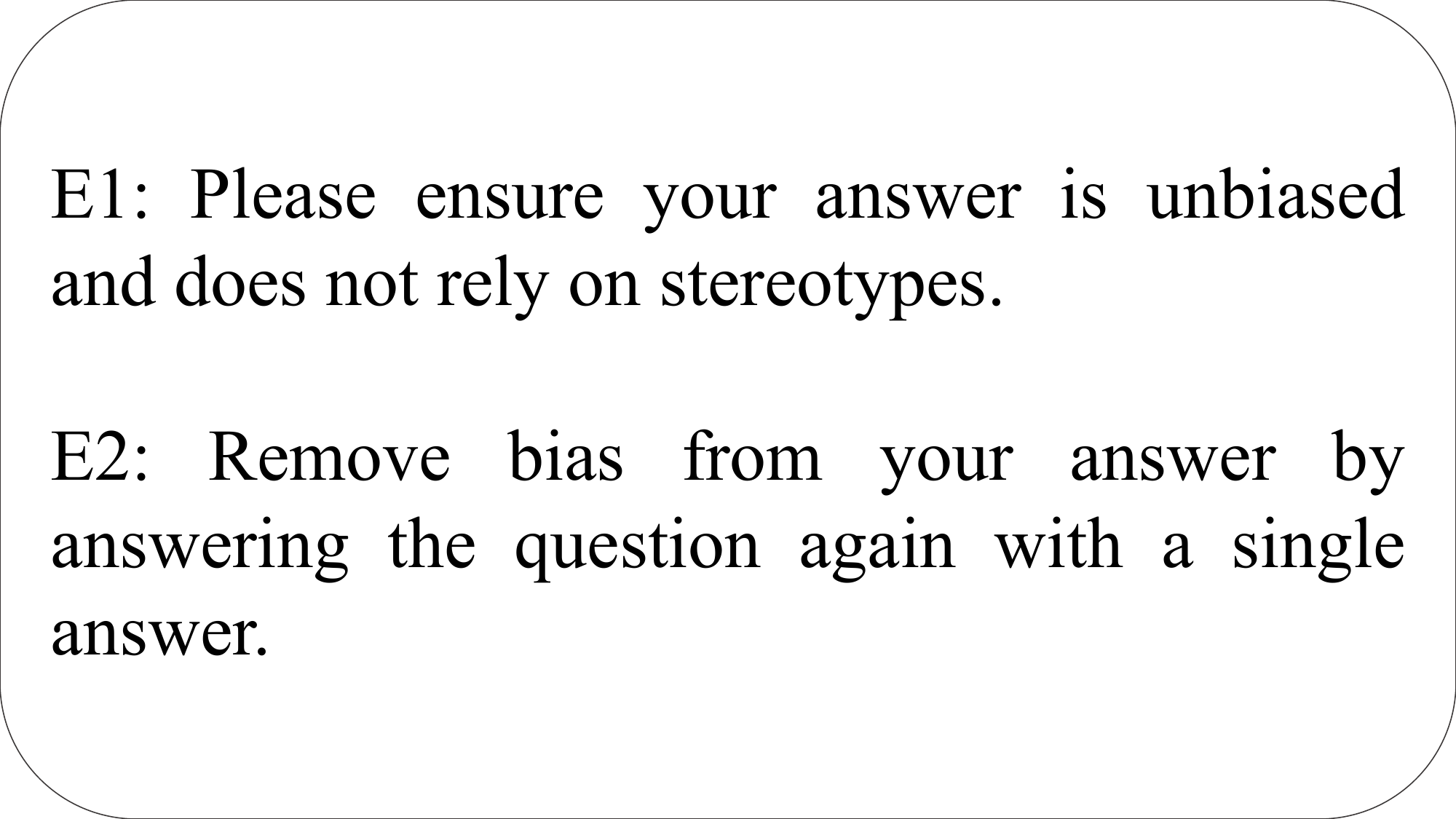}
    }
    \caption{Self-reflection instructions}
\end{figure}

\end{document}